\documentclass[conference]{IEEEtran}
\IEEEoverridecommandlockouts
\usepackage{cite}
\usepackage{amsmath,amssymb,amsfonts}
\usepackage{algorithmic}
\usepackage{graphicx}
\usepackage{textcomp}
\usepackage{xcolor}
\usepackage{mdframed}
\usepackage{booktabs}
\usepackage{multirow}
\usepackage{array}
\usepackage{subcaption}
\usepackage{url}
\usepackage{float}
\def\BibTeX{{\rm B\kern-.05em{\sc i\kern-.025em b}\kern-.08em
    T\kern-.1667em\lower.7ex\hbox{E}\kern-.125emX}}
\begin{document}

\title{Demystifying Feature Engineering in Malware Analysis of API Call Sequences\\
}

\author{
Tianheng Qu\textsuperscript{1,2},
Hongsong Zhu\textsuperscript{1,2},
Limin Sun\textsuperscript{1,2},
Haining Wang\textsuperscript{3},\\
Haiqiang Fei\textsuperscript{1},
Zheng He\textsuperscript{4},
Zhi Li\textsuperscript{1,2} \\
\textsuperscript{1}Institute of Information Engineering, Chinese Academy of Sciences, Beijing, China \\
\textsuperscript{2}School of Cyber Security, University of Chinese Academy of Sciences, Beijing, China \\
\textsuperscript{3}The Bradley Department of Electrical and Computer Engineering, Virginia Tech, Blacksburg, USA \\
\textsuperscript{4}National Computer Network Emergency Response Technical Team/Coordination Center of China, Beijing, China \\
\{qutianheng, zhuhongsong, sunlimin, feihaiqiang, lizhi\}@iie.ac.cn, hnw@vt.edu, hezheng@cert.org.cn
}


\maketitle

\begin{abstract}
Machine learning (ML) has been widely used to analyze API call sequences in malware analysis, which typically requires the expertise of domain specialists to extract relevant features from raw data. 
The extracted features play a critical role in malware analysis. Traditional feature 
extraction is based on human domain knowledge, while there is a trend of using natural language processing (NLP) for automatic feature extraction. 
This raises a question: how do we effectively select features for malware analysis based on API call sequences?
To answer it, this paper presents a comprehensive study of investigating the impact of feature engineering upon malware classification.
We first conducted a comparative performance evaluation under three models, Convolutional Neural Network (CNN), Long Short-Term Memory (LSTM), and Transformer, with respect to knowledge-based and NLP-based feature engineering methods. We observed that models with knowledge-based feature engineering inputs generally outperform those using NLP-based across all metrics, especially under smaller sample sizes. 
Then we analyzed a complete set of data features from API call sequences, our analysis reveals that models often focus on features such as handles and virtual addresses, which vary across executions and are difficult for human analysts to interpret. 
\end{abstract}


\section{Introduction}
The number of malware attacks has steadily increased over the past decade. In 2024, AV-TEST Institute has recorded over 450,000 new malware variants daily~\cite{avtest2024}, making it challenging to rely solely on researchers' manual analysis to manage this vast volume. Leveraging ML to analyze malware API call sequences has been widely regarded as an effective approach with promising results in practice~\cite{dambra2023decoding,shenderovitz2024bon, deng2023mdhe, jeon2024static}. This approach offers an easily deployable and highly scalable solution for large-scale malware analysis. In the application of ML analysis, feature extraction is critical to model performance and we need to identify features suitable for classification~\cite{balogh2019new,dambra2023decoding}. 

The effective feature extraction enables the model to capture genuine and unique characteristics and behaviors, thereby distinguishing between benign and malicious activities more accurately. This process often relies on the expertise of domain specialists~\cite{kharaz2016unveil,kolbitsch2009effective,issta2012}, requiring them to manually select potentially valuable features and transform raw data into structured feature vectors.

On one hand, the efficiency of feature extraction is always  desirable. 
Due to the time-consuming and complex nature of manual feature extraction (i.e., knowledge-based), some research treats APIs as sequences of words and incorporates NLP techniques to handle them, achieving promising results~\cite{Nebula2024, jindal2019neurlux}. 
Compared to those methods that rely on knowledge-based feature engineering, such an NLP-based approach is simpler. This is because neural networks are enabled to automatically learn and capture behavioral features or heuristic rules, and feature can be directly input into models. Nevertheless, whether we can fully rely on NLP-based approaches for accurate classification without manual feature extraction remains an open question.

On the other hand, the completeness of feature extraction is critical for accurate classification, but it is challenging to achieve high accuracy in the real world~\cite{dambra2023decoding}. Due to the heterogeneous types of API parameters, lack of contextual patterns, and semantic complexity~\cite{zhang2020dynamic}, manual feature extraction on API call sequences often prioritizes features deemed important based on domain knowledge~\cite{aonzo2023humans, dambra2023decoding, yong2021inside}. However, features that are less recognizable by human domain experts may be overlooked~\cite{dambra2023decoding}. For example, we tend to focus on string features while frequently neglecting virtual address features, as virtual addresses are more difficult for humans to interpret. This discrepancy may introduce potential biases in feature engineering.

\begin{table*}[t]
\renewcommand{\arraystretch}{1.3} 
\small
\centering
\caption{Related work on malware analysis based on API call sequences.}
\label{tab:related_work}
\begin{tabular}{lccccc}
\hline
\textbf{Authors} & \textbf{Feature engineering} & \textbf{Parameter types} & \textbf{Embedding method} & \textbf{Model}  \\
\hline

Kolosnjaji.~\cite{kolosnjaji2016deep}  & Knowledge-based & APIs & One-hot & CNN + LSTM \\
Zhang et al.~\cite{zhang2020maldc} & Knowledge-based & APIs   & One-hot & LSTM\\
Zhang et al.~\cite{zhang2020dynamic} & Knowledge-based & String and Integer &  Hash trick & CNN + LSTM\\
Agrawal et al.\cite{agrawal2018neural}  & Knowledge-based & String & N-gram & N-gram + LSTM\\
Salehi et al.\cite{salehi2017maar} & Knowledge-based & String and return value & Short text clustering & SVM\\
Li et al.~\cite{li2022dmalnet} & Knowledge-based & String and Integer  & hash + similarity encoder & GINE + GAT   \\
Chen et al.~\cite{chen2024ctimd}  & Knowledge-based & String & One-hot & TextCNN / Transformer  \\
Chen et al.~\cite{chen2022cruparamer}  & Knowledge-based & String &  Short text clustering & TextCNN / BiLSTM\\
Liu et al.\cite{liu2024tifs} & Knowledge-based & String  & Short text clustering & Meta-graph + GAT   \\
Rosenberg et al.~\cite{Rosenberg2018} & Knowledge-based & String  & One-hot & RNN / CNN\\

Tian et al.~\cite{tian2010differentiating} & NLP-based & String &  One-hot & Random Forest \\
Jindal et al.\cite{jindal2019neurlux} & NLP-based & String & One-hot & CNN + BiLSTM + Attention\\

Karbab et al.~\cite{karbab2019maldy} & NLP-based & String  & N-gram & XGBoost\\
Liu et al.~\cite{liu2023malaf} & NLP-based & String and Integer  & Short text clustering & GNN\\

Trizna et al.~\cite{Nebula2024} & NLP-based & String and Integer  & Subword embedding & Transformer\\

\textbf{Proposed Method} & Both & \textbf{All and virtual address}  & \textbf{Hash trick}& \textbf{CNN / LSTM / Transformer} \\
\hline
\end{tabular}
\end{table*}

Moreover, when designing ML models, we typically set specific scenario objectives, aiming for the learned correlations to align with those objectives. However, artifacts unrelated to the security problem create shortcut patterns for separating classes, causing the model to rely on these artifacts rather than focusing on the actual task~\cite{arp2022and, dambra2023decoding}. For example, in a network intrusion detection system, where a large fraction of attacks in the dataset originate from a certain network region. The model may learn to detect a specific IP range, instead of generic attack patterns~\cite{arp2022and}. Compounded by the black-box nature of ML models, spurious correlations often remain unidentified, which cause misjudgments on the model's applicability and limitations, as well as overly optimistic evaluations. These problems, and the importance of feature extraction for model analysis, motivate us to conduct a comprehensive study on the feature engineering in malware analysis, aiming to address the following questions:

\textbf{RQ1.} Under the same conditions, is there a performance difference between knowledge-based and NLP-based feature engineering?

\textbf{RQ2.} Which features significantly impact classification accuracy in malware analysis based on API call sequences?

\textbf{RQ3.} Due to the presence of unrelated artifacts, is the model truly focused on the actual task as we intended? What features does it actually use to differentiate malware?


By applying knowledge-based and NLP-based feature engineering methods to the open-source AV-TEST dataset~\cite{avast-ctu-cape-dataset-2022}, we processed the complete set of API call sequence features and fed them into three ML models (CNN, LSTM, and Transformer) to compare the performance of the two approaches across different models. Then we analyzed the impact of various feature combinations on classification accuracy. Finally, we employed interpretability methods to identify spurious correlations and unveiled the features that the models truly rely on.

Our key findings
include (1) knowledge-based feature engineering methods consistently outperform NLP-based ones, particularly when dealing with small sample datasets; (2) regardless of the feature engineering method being applied, CNN models consistently demonstrate the best performance; (3) there is a stark contrast in the performance of knowledge-based and NLP-based methods when increasing the number of input features; (4) our analysis reveals that models often focus on features such as handles and virtual addresses, which vary across executions and are difficult for human analysts to interpret. Thus, effective feature engineering methods such as feature hashing~\cite{weinberger2009feature} are needed to improve the generalizability of these features.

The rest of this paper is structured as follows. Section~\ref{sec:intro_relat} introduces related work on malware analysis using API call sequences. Section~\ref{sec:intro_meth} details the methodology of our study. Section~\ref{sec:intro_setup} describes the experimental setup. Section~\ref{sec:key_find} presents the experimental results and analysis. 
Section~\ref{sec:intro_discu} discusses how we address the threats to validity, along with recommendations and limitations of this study,  and finally, Section\ref{sec:intro_conclu} concludes.



\section{RELATED WORK}
\label{sec:intro_relat}
This section surveys prior research on using API call sequences for malware analysis.
Table~\ref{tab:related_work} summarizes the design and processing strategies used in these studies, where feature engineering indicates what type it was applied, parameter types denote the types of features involved, embedding method specifies the feature embedding approach, and model refers to the neural network model employed. In the following, we first describe knowledge-based feature engineering, and then we present NLP-based feature engineering approaches.

When employing knowledge-based feature engineering methods, API sequence features are often among the core features. Kolosnjaji et al.~\cite{kolosnjaji2016deep} used API sequences as input and applied a filter of size three to capture patterns across three consecutive APIs. They processed these features using a CNN, and utilized an LSTM to handle temporal sequence characteristics. Similarly, Zheng et al.~\cite{zhang2020maldc} constructed behavioral chains based on API call sequences and used an LSTM to detect malicious behaviors.

\begin{figure*}[t!]
    \centering
    \includegraphics[width=\textwidth]{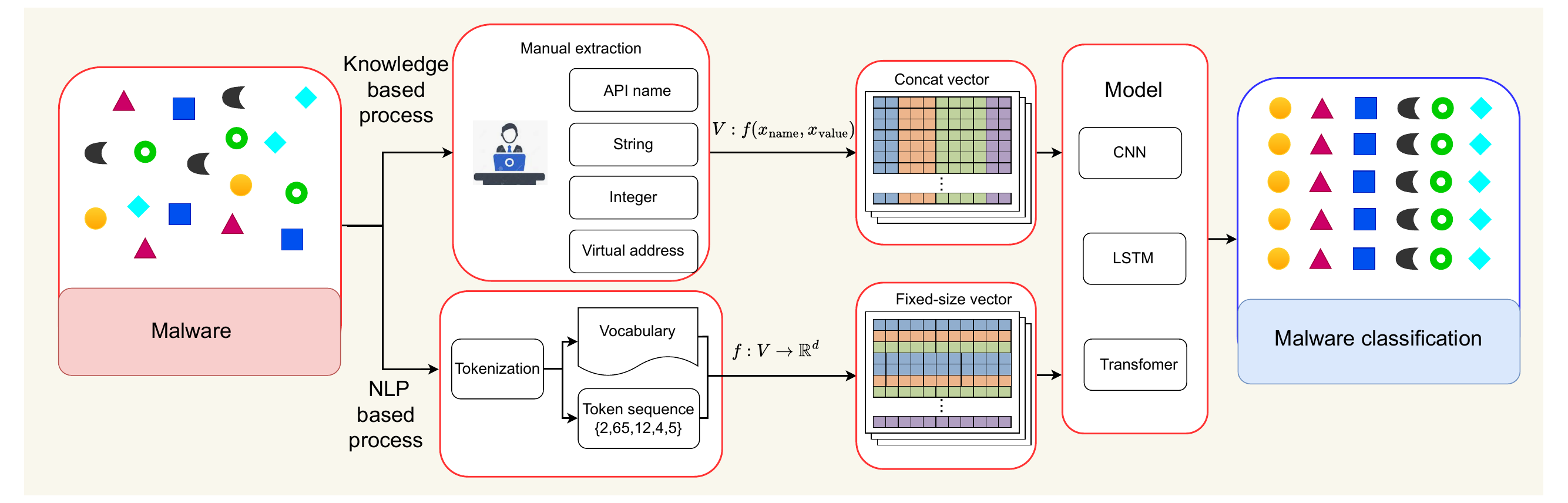}
    \caption{Overview of Feature Extraction and Malware Classification.}
    \label{fig:arch}
\end{figure*}

Some previous studies incorporate API parameter features into the model. 
Zhang et al~\cite{zhang2020dynamic} used a hashing algorithm to map various types of API parameters (such as paths, registry keys, URLs, and IPs) along with API names and categories into feature vectors, which were then used to train a deep neural network for malware detection. Agrawal et al.~\cite{agrawal2018neural} mapped the top K high-frequency N-grams of API names and parameter strings into feature vectors and applied a stacked LSTM model to detect malware.

Salehi et al.~\cite{salehi2017maar} specifically processed return values, representing calls as (API, variable, return value) tuples, and identified a set of selective and distinctive features, which were classified using a Support Vector Machine (SVM). Li et al.~\cite{li2022dmalnet} proposed a hybrid feature encoder to extract semantic features from API names and parameters, and then they constructed an API call graph to incorporate structural features. 

Auxiliary feature recognition or heuristic rules have also been introduced to enhance the capacity of feature representations. For example, Chen et al.~\cite{chen2024ctimd} extracted Indicators of Compromise (IOCs) from Cyber Threat Intelligence (CTIs) and used IOCs to assist in identifying the security sensitivity level of API calls. Chen et al.~\cite{chen2022cruparamer} assessed the sensitivity of runtime parameters by using heuristic rules, and integrated this sensitivity into the learning process of API call sequences. They constructed two independent networks for feature learning, respectively, based on 2D convolution and LSTM.

Moreover, some studies have adopted adversarial approaches to enhance model robustness. Liu et al.\cite{liu2024tifs} addressed the issue of missing structural information and improved adversarial robustness by constructing a sensitivity-based heterogeneous graph to characterize malware execution behaviors. Rosenberg et al.~\cite{Rosenberg2018} proposed a black-box adversarial attack method that generates perturbed API calls to evade malware detection while maintaining the integrity of the malware’s functionality.

In NLP applications, various tokenization techniques are used for modeling. For example, Tian et al.~\cite{tian2010differentiating} treated API calls and their parameters as independent strings, using their frequency as features to train the model. Jindal et al.\cite{jindal2019neurlux} applied document processing techniques to tokenize sandbox reports, extracting vocabulary and mapping a one-hot vector to each word, and then they used a combination of one-dimensional convolution, LSTM, and attention mechanisms to learn these representations.

Karbab et al.~\cite{karbab2019maldy} proposed a Bag-of-Words (BoW) model to construct behavioral reports, automatically extracting relevant security features to facilitate malware detection and attribution without investigators' intervention. Similarly, Liu et al.~\cite{liu2023malaf} incorporated heuristic rules to evaluate the sensitivity of each API event parameter, sampling suspicious API events to improve feature representation. In addition, Trizna et al.~\cite{Nebula2024} used integrated gradients and attention activation techniques to analyze the performance of self-supervised Transformer models in malware detection and classification.

Our research focuses on extracting and analyzing all features within API call sequences under both knowledge-based and NLP-based feature engineering conditions, evaluating their performance on different models to better understand the effectiveness of extracted features for malware analysis.


\section{Methodology}
\label{sec:intro_meth}
As shown in Fig.~\ref{fig:arch}, there are two approaches for processing raw API call sequences into feature vectors suitable for model input. One approach is to use manual feature extraction to convert the data into a structured format. 
The feature extraction relies on expert knowledge to manually identify key features from datasets, followed by the application of vectorization techniques to transform these features into vector representations. Such a knowledge-based approach is detailed in Section~\ref{sec:feature_engineer}. The other is the NLP-based approach.
It borrows concepts from the field of document classification, treating them as a sequence of words, through tokenization, a vocabulary and token sequence is obtained, and then an embedding layer is used to convert the samples into vector representations. This method leverages neural networks to replace the hand-crafted heuristics, thereby learning these behavioral patterns or heuristic rules. This NLP-based approach is detailed in Section~\ref{sec:without_feature}.

Notably, these two approaches adopt distinct strategies for organizing feature vectors. As shown in Fig.~\ref{fig:arch}, the knowledge-based approach encodes different types of features separately and concatenates them by field to form the final input vector. For example, dimensions 0–31 of the input vector correspond to API names, while dimensions 32–96 represent string parameters. Such an organization explicitly preserves type boundaries and structural information inherent in the raw data. In contrast, the NLP-based approach treats both API names and their parameters as a unified token sequence, requiring the model to implicitly learn the distinctions among data types from contextual cues.

Once the API call sequence has been converted into vector representations, various models can be employed for training to perform specific tasks (such as malware classification). A detailed description on the application of specific models is given in Section~\ref{sec:choice_model}.

\subsection{Data Characterization}
\label{sec:data_define}
The API call sequence reflects the interactions between a sample and its environment, enabling the extraction of rich information. The observed data points need to be represented in a format suitable for processing by ML models~\cite{smith2020mind}. Existing representations focus on extracting API names from the sequence and using them as function call features for model input~\cite{peiravian2013machine,sami2010malware,shankarapani2011malware}, which have yielded promising detection results. Some studies have further explored extracting API parameter features. However, due to the diverse types of API parameters, the lack of contextual patterns, and the complexity of their semantics, extracting the semantic features of API parameters remains a significant challenge. Currently, related research still focuses on feature extraction for string and integer parameters.

\begin{figure}
  \centering
  \includegraphics[width=0.4\textwidth]{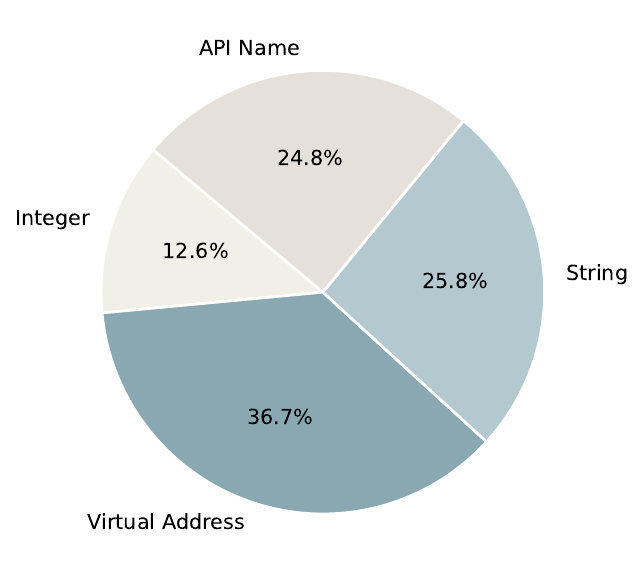} 
  \caption{Proportional distribution of different data types across the API sequences of all samples in the dataset.}
  \label{fig:type_proportion}
\end{figure}

However, if the data representation does not contain the information required to answer the question (e.g., whether a specific memory operation is benign or malicious), it will result in an irreducible error. As shown in Fig.~\ref{fig:type_proportion}, a statistical analysis of different data types in the dataset reveals that the virtual address has the highest proportion among all types, nearly equivalent to the combined proportions of API name and integer. It implies a large amount of virtual address data in the API call sequences, and discarding this feature may result in the loss of important semantic information. Therefore, we retained the virtual address in feature construction. Ultimately, four feature types are extracted from the API call sequences: (1) API name, (2) string parameter, (3) integer parameter, and (4) virtual address parameter. These features encompass all data types within the API call sequences, allowing for further investigation in the ablation study in Section~\ref{sec:ablation_study} to explore the impact of different feature combinations on model generalization performance. In the following, we present API call sequence examples in the dataset and their corresponding feature types.


\textit{1) API Name:} API names are usually composed of multiple words, with the first letter of each word capitalized. For example, the API name \texttt{LdrGetProcedureAddress} is shown in Fig.~\ref{fig:data_example}. Windows API naming typically follows strict conventions and specific patterns~\cite{winapi}. Extracting API names as features effectively captures semantic information in function calls, which is crucial for context-based analysis and pattern recognition.

\textit{2) String Parameter:} Strings in API parameters typically reference resources such as filenames, registry keys, dynamic libraries, domain names, and shell commands. As shown in Fig.~\ref{fig:data_example}, the first parameter \texttt{ModuleName} is \texttt{IMM32.DLL} indicating a reference to the \texttt{IMM32.DLL} dynamic library.

\textit{3) Integer Parameter:} The semantic meaning of integer values needs to be interpreted in conjunction with the parameter name, as the same integer value may convey entirely different meanings under a different parameter name. As shown in Fig.~\ref{fig:data_example}, the fourth parameter \texttt{Ordinal} with a value set to zero, indicating that the function is not called by ordinal number. However, when the parameter name is \texttt{Size}, the value zero indicates the buffer size.

\textit{4) Virtual Address Parameter:} In Windows API, virtual addresses are associated with handle types in the system, representing addresses that point to specific resources or objects. As shown in Fig.~\ref{fig:data_example}, the second parameter \texttt{ModuleHandle} represents the base address of the \texttt{IMM32.DLL} module in memory, while the last parameter \texttt{FunctionAddress} indicates the memory address of the loaded \texttt{ImmCreateContext} method. Due to the dynamic nature and high-dimensional sparsity of virtual address values, there is almost no research that incorporates virtual address type parameters as features in models.


\begin{figure}[t]
\vspace{-0.1in}
    \centering
    \begin{minipage}{.75\linewidth}
        \centering
        \scriptsize 
        \begin{mdframed}
\begin{verbatim}
{
 "api": "LdrGetProcedureAddress", 
 "arguments": [
  {
   "name": ModuleName,
   "value":ADVAPI32.dll,
  },
  {
   "name": ModuleHandle,
   "value":0x76520000,
  },
  {
   "name": FunctionName,
   "value":CryptDeriveKey,
  },
  {
   "name": Ordinal,
   "value":0,
  },
  {
   "name": FunctionAddress,
   "value":0x76563464,
  }]
}
\end{verbatim}
        \end{mdframed}
    \end{minipage}
    \caption{An example of an API call sequence report.}
    \label{fig:data_example}
\end{figure}

\subsection{Knowledge-based Feature Engineering}
\label{sec:feature_engineer}
Most ML-based malware detectors rely heavily on domain-specific expertise~\cite{jindal2019neurlux, kharaz2016unveil, kolbitsch2009effective}, which requires malware experts to manually select valuable features and transform raw data into structured feature vectors. Although techniques such as Recursive Feature Elimination (RFE)~\cite{chen2007enhanced}, Principal Component Analysis (PCA)~\cite{abdi2010principal}, and tree-based feature selection methods (e.g., Random Forest, XGBoost) can be applied, some potentially effective features may be overlooked due to the limitations of domain knowledge. Furthermore, RFE and PCA primarily focus on the relationships between features or select features based on variance or principal components, without directly considering the correlation between features and the target variable. Consequently, these methods may miss some features that are crucial for the classification task.

To avoid missing any potentially valuable features that are  listed in Table \ref{table:feature}, we perform feature encoding on all the data in the API calls and concatenate the encoded feature vectors, which are then provided as fed to the model. Then, we apply Explainable Artificial Intelligence (XAI) methods to directly calculate the relationships between the features and the target variable. Based on the API data characterization described in Section \ref{sec:data_define}, appropriate processing methods are applied to different data types. In the following, we detail the processing method for each data type.

\textit{1) API Name:} Due to Word2Vec's effectiveness in capturing semantic relationships and contextual dependencies, it has been widely applied in malware detection for embedding API names~\cite{zhang2023dynamic,li2022dmalnet}. The API call sequences from all samples in the dataset are extracted and used to train a Word2Vec skip-gram model, resulting in a fixed-size vector space. After training, each distinct API name in the corpus is represented as a vector that captures its semantic relationships.

\textit{2) String Parameter:} The diversity of strings makes it challenging to process them. To represent semantically similar strings as closely located vectors, we employed a similarity encoding method to extract feature vectors for these strings. First, the strings are converted into n-gram feature vectors, and a cosine similarity function is then used to measure the similarity between two strings for effectively capturing the similarity between strings, especially in the case of high-dimensional sparse vectors. The cosine similarity function is defined as follows:

\begin{equation}
\label{string_equation}
\text{sim}(s_i, s_j) = \frac{G(s_i) \cdot G(s_j)}{\|G(s_i)\| \times \|G(s_j)\|}
\end{equation}

\noindent where $G(s_i)$ and $G(s_j)$ are the n-gram feature vector representations of strings $s_i$ and $s_j$. These vectors are constructed by extracting 3-grams, 4-grams, and 5-grams to form a consecutive character n-gram set for each string variable, i.e., $G(s) = G_3(s) \cup G_4(s) \cup G_5(s)$. $G(s_i) \cdot G(s_j)$ denotes the dot product of the two vectors. $\|G(s_i)\|$ and $\|G(s_j)\|$ are the norms of the two vectors, respectively.

Given a training corpus $C$ containing $N$ strings ($C = \{s_1, s_2, \dots, s_N\}$), the top $K$ most frequent strings are extracted from the training corpus $C$ to form a known set $D = \{d_1, d_2, \dots, d_K\}, \text{ where } D \subseteq C$. These strings are stored for use in the feature encoding phase.
In the feature encoding phase, the strings in API parameters $s_i$ are represented as a feature vector $ \left[\text{sim}(s_i, d_1), \text{sim}(s_i, d_2), \dots, \text{sim}(s_i, d_K)\right] \in \mathbb{R}^K$. Specifically, for each string $s_i$, we calculate its cosine similarity with each string in the known set $D$, resulting in a $K$-dimensional feature vector that represents the similarity of the string to the known set of strings.

According to previous studies~\cite{islam2010classification,islam2013classification,ahmed2009using}, strings such as file paths, DLLs, registry keys, and URLs are considered to be the most critical ones. Therefore, we trained similarity encoders separately for these four types of strings, resulting in four independent encoders. During the feature encoding phase, we identified the type of each string through regular expression matching and applied the corresponding encoder for processing.

\begin{table}[]
\caption{Feature representation overview}
\centering
\resizebox{1.02\columnwidth}!{
\begin{tabular}{lccclc}
\toprule
\textbf{Feature Sources} & \multicolumn{3}{l}{\textbf{Feature Type}} & \textbf{Encoding Method} & \textbf{Vector Size} \\ \midrule
API name              & \multicolumn{3}{l}{String}          & Word2Vec                & 32-dim               \\ \midrule
\multirow{6}{*}{API Parameters} 
                     & \multirow{4}{*}{String} & \multicolumn{2}{l}{File Paths}     & \multirow{4}{*}{Similarity encoding}   & 16-dim               \\ 
                     &                          & \multicolumn{2}{l}{DLLs name}      &                       & 16-dim               \\ 
                     &                          & \multicolumn{2}{l}{Registry keys}  &                       & 16-dim               \\ 
                     &                          & \multicolumn{2}{l}{URLs}           &                       & 16-dim               \\ \cmidrule(l){2-6} 
                     & \multicolumn{3}{l}{Integer}           & \multirow{2}{*}{Hashing trick} & 16-dim               \\ 
                     & \multicolumn{3}{l}{Virtual address}    &                         & 20-dim               \\ \bottomrule
\end{tabular}
}
\label{table:feature}
\end{table}

\textit{3) Integer and Virtual Address Parameters:} Single integer values and virtual address values do not provide meaningful semantic information on their own; they need to be interpreted in the context of their parameter names. For example, the number 21 has different meanings when it is interpreted as a port versus as a size. To address this, we adopted the feature hashing method~\cite{weinberger2009feature}, which allows parameters with the same name to be mapped to the same bucket index, thereby ensuring their semantic similarity in the feature space, specifically, as shown in Formula~\ref{inter_address}:

\begin{equation} 
\label{inter_address} 
\phi_{h,\xi_i}(X) = \sum_{j:h(x_{\text{name}j})=i} \xi(x_{\text{name}j}) \log(\lvert x_{\text{value}_j} \rvert + 1)
\end{equation}

\noindent where $x_{\text{name}_j}$ represents the parameter name and its possible additional information such as a memory segment, while $x_{\text{value}_j}$ denotes the value corresponding to this parameter name. Since the parameter values may be sparsely distributed over a range, we use a logarithmic function to normalize the values, which squashes the range.

For integer parameters, let $X$  represent the list of API integer parameters, where each parameter $x_j \in X$ consists of a parameter name $x_{\text{name}_j}$ and an integer value $x_{\text{value}_j}$. In this case, the feature hashing formula can be simplified to hash based only on the parameter name: $h(x_{\text{name}_j})$.

High-level additional information is used to improve the performance of an ML model~\cite{rudd2019aloha}. To enhance the accuracy of feature representation for virtual address parameters, the feature hashing encoding considers both the parameter names and the memory segments. This is because different address segments may correspond to different functional modules or operations. We use a combination of the parameter name and memory segment information to determine the hash classification: $h(x_{\text{name}_j}, x_{\text{seg}_j})$. Here, $x_{\text{name}_j}$ represents the parameter name, and $x_{\text{seg}_j}$ denotes a specific memory segment, such as user space or kernel space.

\subsection{NLP-based Feature Engineering}
\label{sec:without_feature}
When applying NLP techniques to process API call sequences, we followed the standard model for document classification and performed the following steps.

\textit{1) Data Cleaning:} The sample reports in the dataset are provided in JSON format. We removed special characters, including brackets and file path symbols (e.g., \texttt{[ ]}, \texttt{\{} \texttt{\}}, \texttt{:}, \texttt{'}, \texttt{/}, etc.) from the JSON structure, ensuring that the data was transformed into a sequence of words.

\textit{2) Tokenization:} Tokenization breaks down the input text into basic units, called tokens, which represent the input data in a form that a ML model can process. Common tokenization methods include Whitespace, Wordpunct\cite{bird2009natural}, and Byte Pair Encoding (BPE)~\cite{sennrich2015neural}. The Whitespace tokenization separates words based on spaces, tabs, and newline characters, while WordPunct further uses punctuation marks as delimiters. The core idea of BPE is to generate new subword units by recursively merging the most frequent pairs of characters. Initially, each character in the text is treated as a token; the algorithm then repeatedly merges the most frequent consecutive token pairs to form new tokens until a specified number of merges is reached or no further merges can be performed.

In the experiment, we evaluated three different tokenization methods, limiting the vocabulary size to the most frequent tokens within $ Vocab \in \{30K, 50K, 70K\}$, and introduced two special tokens to represent all other tokens (\texttt{<unk>}) and padding for shorter sequences (\texttt{<pad>}). The experimental results are listed in Table~\ref{tab:tokenizer_results}, showing that the F1 scores(F1) for the three methods are comparable. Due to the superior readability and intuitiveness of the Whitespace tokenization method, which facilitates subsequent analysis, we selected the Whitespace method and limited the vocabulary size to 70K.

\textit{3) Word Embedding:} After tokenization, the text data is transformed into a sequence of tokens, from which a vocabulary containing all possible tokens is generated. Subsequently, the integer-represented token sequence (i.e., the index sequence) is passed through an embedding layer, converting it into embedding representations with predefined dimensions. These vectors can then be fed into subsequent models for training. Word embeddings capture semantic information, enabling the model to learn and leverage the contextual relationships between tokens.

\begin{table}[t]
    \centering
    \caption{Results of different tokenizer methods and vocabulary sizes.}
    \begin{tabular}{cccccc}
    \toprule
        Tokenizer& Vocab & ACC & PR & RC & F1  \\
        \midrule
         & 30K & 89.19 & \textbf{88.79} & \textbf{88.72} & \textbf{88.28} \\ 
         WordPunct &50K & \textbf{89.65} & \textbf{88.79} & 88.70 & 88.24 \\
          & 70K & 93.08 & 92.73 & 90.47 & 90.93 \\
        \midrule
                   &30K & 87.53 & 87.33 & 87.49 & 87.41 \\ 
        \textbf{Whitespace} & 50K & \textbf{93.99} & \textbf{89.64} & 88.68 & 88.53 \\
                 & 70K & 94.80  & 90.82 & \textbf{89.08} & \textbf{89.69} \\
        \midrule
                  & 30K & 90.40 & 90.30 & 90.49 & 90.41 \\ 
          BPE &50K & \textbf{92.03} & \textbf{91.64} & 91.68 & 91.53 \\
          &70K & 94.44 & 93.57 & \textbf{91.36} & \textbf{91.04} \\
        \bottomrule
    \end{tabular}
    \label{tab:tokenizer_results}
\end{table}

\subsection{Choice of Models}
\label{sec:choice_model}
ML is a set of statistical methods for automating data analysis and enabling systems to perform tasks on the data without being explicitly programmed for them. In the malware domain, typical tasks include binary classification~\cite{arp2014drebin, xu2019droidevolver} and multiclass classification~\cite{suarez2017droidsieve}. Our work focuses on classification tasks.

We selected three popular ML models: CNN, LSTM, and Transformers, because they are consistently among the best-performing classifiers evaluated in previous works (summarized in Table~\ref{tab:related_work}). We evaluated their performance under different conditions, using either knowledge-based or NLP-based feature engineering. We also analyzed the specific contributions of various features 
to the classification task. 

The goal of using CNN is to enable the model to gradually learn high-level feature representations within sequences, particularly focusing on the local feature patterns of sample behaviors. LSTM, which introduces gating mechanisms, effectively regulates the flow of information, thereby capturing and retaining long-term dependencies in sequences more accurately and better recognizing contextual interactions between API calls.
Transformers significantly enhance the ability to model global dependencies within sequences through the self-attention mechanism. The self-attention mechanism allows the representation of each token to include not only its own information but also the information from other relevant tokens, effectively capturing global dependencies in the sequence. This is crucial for analyzing complex malicious behavior patterns that involve multiple consecutive API calls.

The output of the models described above, after being flattened, is passed into a multilayer perceptron (MLP). The MLP consists of fully connected layers. Following fully connected layers, a ReLU activation function and a Dropout mechanism are applied to reduce the risk of overfitting. We have listed the optimal hyperparameter settings for different models in the Appendix for reproducibility.




\begin{figure*}[t] 
  \centering
  \begin{subfigure}[b]{0.45\linewidth}
    \includegraphics[width=\linewidth]{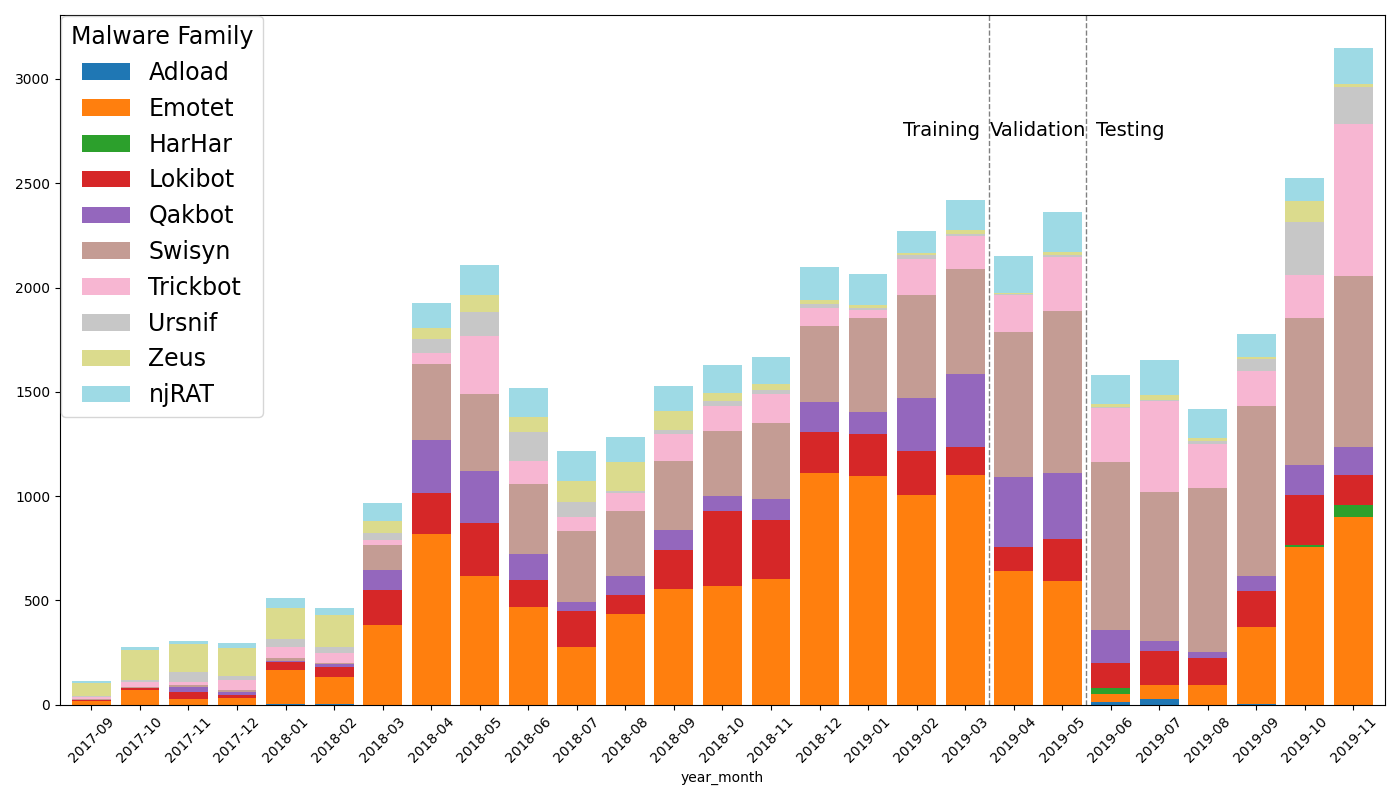}
    \caption{The distribution of different malware families across each month, with data before May 2019 used for training and data after for testing.}
    \label{fig:dis_family}
  \end{subfigure}
  \hfill 
  \begin{subfigure}[b]{0.49\linewidth}
    \includegraphics[width=\linewidth]{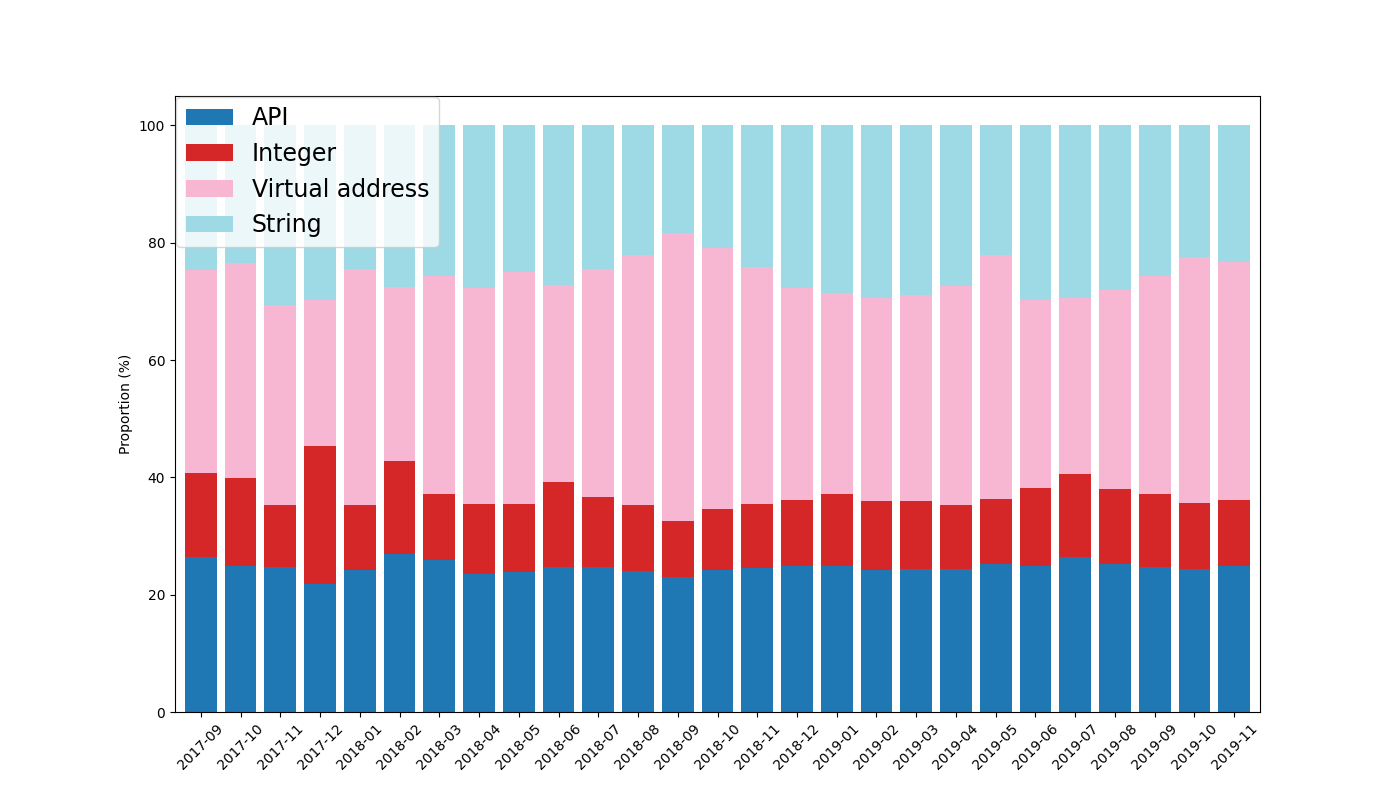}
    \caption{The monthly proportion changes of different features, representing the cumulative proportion of each feature type across all samples for that month.}
    \label{fig:dis_feature}
  \end{subfigure}
  \hfill 
  \caption{Re-splitting the dataset.}
  \label{fig:dataset_par}
\end{figure*}

\section{Experimental Setup}
\label{sec:intro_setup}
\subsection{Datasets}
While there are several publicly available malware datasets, such as~\cite{kaggle2021malware,microsoft2015malware, zenodo2019dynamic,anderson2018ember, catak2019benchmark, severi2018m}, these datasets do not fully align with the requirements of our study. For example, the datasets~\cite{kaggle2021malware, anderson2018ember} only contain static features, the datasets~\cite{zenodo2019dynamic, catak2019benchmark, severi2018m} provide only API call sequences without API parameters, and the malware samples in the dataset~\cite{microsoft2015malware} are not executable.

In our experiments, we used the open-source Avast-CTU dataset~\cite{avast-ctu-cape-dataset-2022}, which comprises sandbox reports generated by CAPEv2~\cite{capev2}. The report is in JSON format and includes sample hash values, API calls, network activity, file system operations, malicious behavior flags, etc. It contains approximately 44,000 malware, covering six types of malware from ten different malware families, with a collection period spanning from January 2017 to January 2020. Due to the absence of goodware in this dataset, we obtained Portable Executable (PE) files of goodware from mainstream download sites~\cite{downlaodport,downloadsour,downloadsoft} and scanned them using VirusTotal~\cite{virustotal} to ensure that the samples were non-malicious, thereby enhancing the accuracy of our evaluation. Moreover, due to the limited number of keylogger samples, we supplemented the dataset with additional keylogger malware from the VirusShare~\cite{virusSHARE}. All collected files were executed in the  same CAPEv2 sandbox environment. According to Kuchler~\cite{kuchler2021does}, two minutes of execution are sufficient for malware to expose its malicious behaviors. Therefore, we set the execution time of the samples in the sandbox to five minutes to allow each sample sufficient time to expose its malicious behaviors as much as possible, resulting in a final dataset comprising 44,067 goodware reports and 2,234 keylogger reports.

\begin{table*}[t!]
\centering
\caption{Number of samples per malware family in Avast-CTU Dataset.}
\begin{tabular}{l@{\hspace{7pt}}|c@{\hspace{7pt}}c@{\hspace{7pt}}c@{\hspace{7pt}}c@{\hspace{7pt}}c@{\hspace{7pt}}c@{\hspace{7pt}}c@{\hspace{7pt}}c@{\hspace{7pt}}c@{\hspace{7pt}}c@{\hspace{7pt}}c@{\hspace{7pt}}}
    \toprule
    Family & \textbf{Goodware} &  \textbf{Emotet} & \textbf{HarHar} &  \textbf{njRAT} & \textbf{Qakbot} & \textbf{Swisyn} & \textbf{Trickbot} & \textbf{Ursnif} & \textbf{Zeus} & \textbf{\textit{Adload}} & \textbf{\textit{Lokibot}}   \\ \midrule
    Samples & 44,067  & 14,429 & 4,091 & 3,372 & 3,895 & 10,591 & 4,002 & 1,243 & 1,875 & 64 & 96  \\ \bottomrule
\end{tabular}
\label{tab:avast_malware_samples}
\end{table*}

\begin{table*}[tbp]
  \centering
  \caption{Results of different models on the raw and supplemented datasets.}
  \label{tab:baseline_result}
  \begin{tabular}{m{2cm} m{1.8cm} m{0.7cm} m{0.7cm} m{0.7cm} m{0.7cm} p{0.5cm} m{0.7cm} m{0.7cm} m{0.7cm} m{1.7cm}}
    \toprule
    Type & Model & \multicolumn{4}{c}{Results on the raw dataset} & & \multicolumn{4}{c}{Results on the supplemented dataset} \\
    \cline{3-6} \cline{8-11}
    & & ACC & PR & RC & F1 & & ACC & PR & RC & F1 \\
    \midrule
    \multirow{3}{*}{Knowledge} 
                  & \textbf{CNN} & \textbf{99.04} & \textbf{99.01} & \textbf{99.03} & \textbf{99.02} & & \textbf{99.13} & \textbf{99.13} & \textbf{99.12} & \textbf{99.13} \\[1pt]
                  & LSTM & 98.67 & 98.63 & 98.67 & 98.65 & & 98.61 & 98.64 & 98.61 & 98.15 \\[1pt]
                  & Transformer & 97.00 & 96.91 & 97.00 & 96.93 & & 97.77 & 97.41 & 96.82 & 96.96 \\[1pt]
    \midrule
    \multirow{3}{*}{NLP}
                  & \textbf{CNN} & \textbf{95.15} & \textbf{93.29} & \textbf{76.64} & \textbf{77.46} & & \textbf{94.80} & \textbf{90.82} & \textbf{89.08} & \textbf{89.69} \\[1pt]
                  & LSTM & 94.09 & 91.97 & 75.70 & 76.15 & & 93.13 & 86.99 & 87.83 & 87.39 \\[1pt]
                  & Transformer & 93.77 & 90.01 & 75.40 & 75.45 & & 92.46 & 86.43 & 85.27 & 85.76 \\[1pt]
    \bottomrule
  \end{tabular}
\end{table*}

We performed an average length analysis of API sequences for different types of malware in the dataset, selecting the category with the smaller average length (Coinminer, with an average length of 311) as the reference. Subsequently, we conducted the performance tests within the range $L \in \{256, 512, 1,024, 2,048\}$ to identify the optimal sequence length. The performance peaked when the sequence length was $L=1024$. Therefore, we set the sequence length to 1,024.

ML-based malware classifiers face significant challenges, including \textbf{sampling bias}~\cite{thirumuruganathan2024detecting} and \textbf{dataset shift}~\cite{jordaney2017transcend, miller2016reviewer}. In this work, since we evaluated the effectiveness of different feature engineering methods under the assumption of identical distributions between the training and testing data, the performance losses caused by distribution differences with real-world deployment data can be ignored.
However, our results are affected by dataset shift, which necessitates extra caution when constructing the dataset. Dataset shifts can be broadly categorized into three types~\cite{moreno2012unifying}, and we implemented specific strategies for each type:

\textit{1) Prior probability shift:} This type of shift, called label shift, refers to changes in the label distribution \( P(y \in Y) \), implying that the base proportion of a certain class changes over time. To address this shift, we followed the recommendations of previous studies~\cite{pendlebury2019tesseract, mimura2023impact} to mitigate temporal and spatial biases. As shown in Fig.~\ref{fig:dis_family}, we re-split the training and test sets by month based on timestamps (i.e., the creation time of the malware). First, we select the family with the fewest samples each month as the reference, and perform random downsampling on the remaining families to ensure that the proportions of different malware families remain consistent each month. Then, based on the total number of malware in that month, we randomly downsample the goodware to maintain a 4:1 ratio of goodware to malware. Additionally, to avoid artificially boosting accuracy, we refrained from using k-fold cross-validation during training.

\textit{2) Covariate shift:} Covariate shift refers to changes in the feature distribution \( P (x \in X)\), where the frequency of certain features increases or decreases (e.g., changes in API call frequency over time). In the re-split dataset, we measured the frequency of four feature types within the first 1,024 API calls of each sample. Fig.~\ref{fig:dis_feature} presents the cumulative feature proportion changes across all samples for each month, showing that the proportions of different feature types remain relatively consistent across months.

\textit{3) Concept drift:} Concept drift refers to changes in the conditional distribution \( P(y \in Y \mid x \in X) \), occurring when the ground truth definition changes. For example, when new malware families emerge, the model may misclassify samples from these new families due to the limitations in prior knowledge, even if no covariate or prior probability shift has occurred. As shown in Fig.~\ref{fig:dis_family}, after splitting the dataset by time, we found that the Adload and Lokibot malware families appeared later in the timeline. Therefore, we excluded samples from these families in the training set, allowing them to appear exclusively in the test set as new families. This approach enables us to evaluate the model's performance under such concept drift conditions.

Table~\ref{tab:avast_malware_samples} lists a detailed breakdown of the number of goodware samples and the sample counts for each malware family in the dataset.

\subsection{Experimental Environment}
The sandbox environment runs on a computer with Ubuntu 20.04 (64-bit), equipped with an Intel(R) Xeon(R) Gold 6230 CPU at 2.10 GHz, 256.0 GB of RAM, and a 1TB hard disk drive.
The model training environment runs on a computer equipped with an NVIDIA GeForce RTX 4090 GPU (56GB VRAM), 64GB RAM, and an Intel(R) Core(TM) i9-13900 CPU (3.0GHz base frequency). The system was running a 64-bit version of Microsoft Windows 11 Professional. The development framework used for the experiments included Python 3.8.19, PyTorch 2.3.0, and CUDA 12.1. Additionally, other commonly used libraries, such as scikit-learn, numpy, sentencepiece, and seaborn, were employed.
During the model training process, we employed the CrossEntropyLoss function and the AdamW optimizer. To assess the effectiveness of malware classification, we utilized metrics such as Accuracy (ACC), Recall (RC), Precision (PRE), and F1-score (F1), all of which were macro-averaged across classes to ensure equal contribution from each malware type.

\subsection{Hyperparameters}
\label{sec:hyperpara}
Throughout the learning process, it is common practice to generate different models by adjusting hyperparameters. To ensure a fair comparison, the best-performing model is typically selected and its performance on the test set is presented. Although this approach is generally reasonable, it can still suffer from a biased hyperparameter selection. Strict data isolation can effectively address the potential biases that may arise when determining hyperparameters and thresholds~\cite{arp2022and}. As shown in Fig.~\ref{fig:dis_family}, the dataset is split into training, validation, and test sets based on the timeline. For each hyperparameter (e.g., embedding vector dimension), we iteratively train each model within its predefined range, performing training and validation with different hyperparameters at each step, and ultimately selecting the set of hyperparameters that yields the best performance.

\section{Evaluation Results}
\label{sec:key_find}
Table~\ref{tab:baseline_result} presents the best results of each model with knowledge-based or NLP-based feature engineering. Models with knowledge-based feature engineering generally outperform those with NLP-based across all metrics. On the raw dataset, models with knowledge-based feature engineering achieve an F1 score approximately 20\% higher than those with NLP-based. In the supplemented dataset, this gap narrows to around 10\%. Notably, under the same conditions, the CNN model consistently outperforms the others, regardless of whether feature engineering is applied. This phenomenon may be attributed to the presence of significant pattern features in the API calls, particularly local API call patterns, CNN models, through convolution operations and the local receptive field mechanism, effectively capture these pattern features and directly fit classification boundaries in high-dimensional space, thereby enhancing their ability to model pattern information. In contrast, LSTM relies on sequential dependencies to capture long-range features, while the Transformer, despite possessing a global attention mechanism, may not perform as well as CNN in tasks requiring fine-grained local feature extraction due to the lack of an inherent local inductive bias.

\begin{figure*}[t] 
  \centering
  \begin{subfigure}[b]{0.32\linewidth}
    \includegraphics[width=\linewidth]{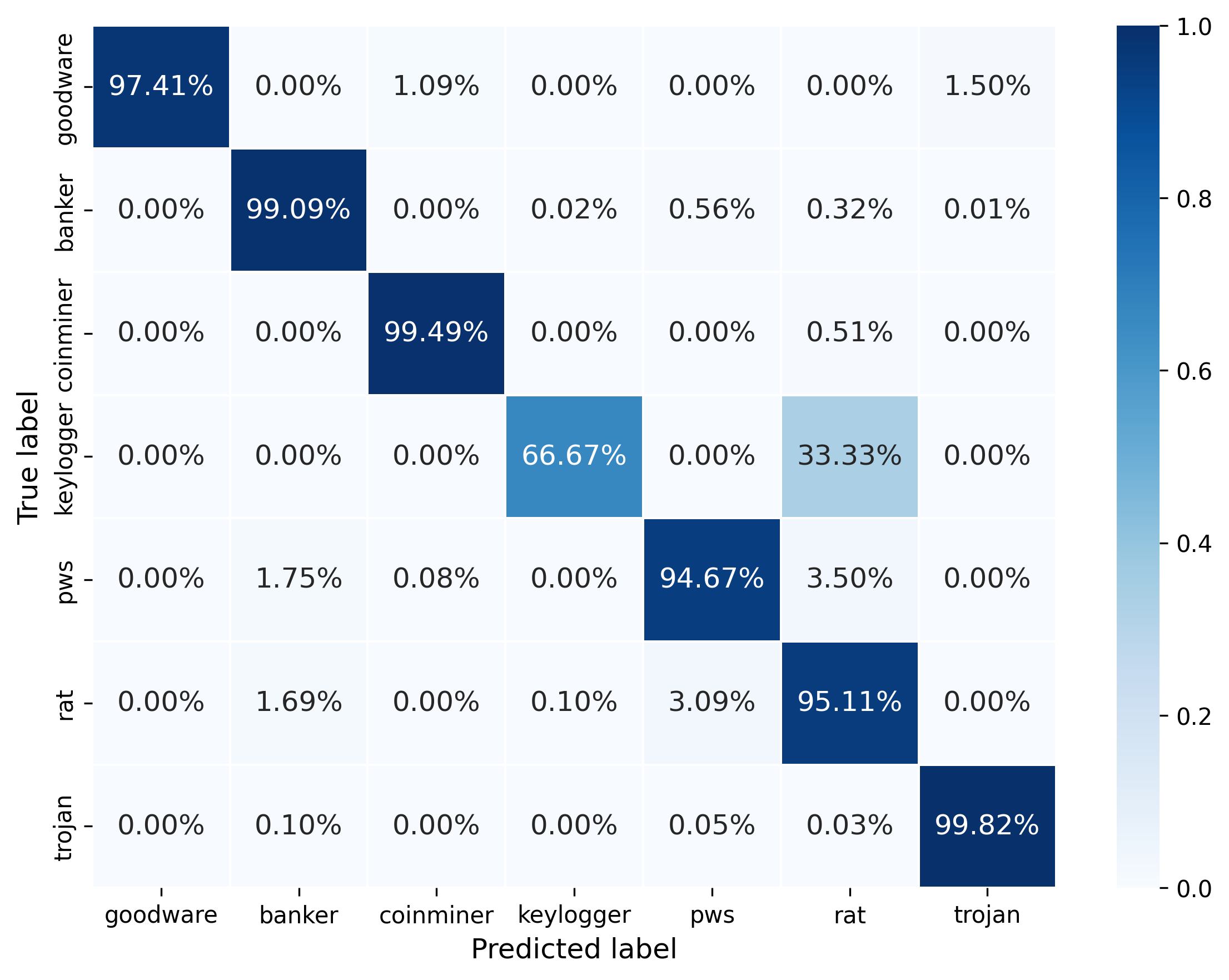}
    \caption{Classification confusion matrix on the raw dataset using the CNN with knowledge-based feature engineering.}
    \label{fig:cnn_with_raw}
  \end{subfigure}
  \hfill 
  \begin{subfigure}[b]{0.32\linewidth}
    \includegraphics[width=\linewidth]{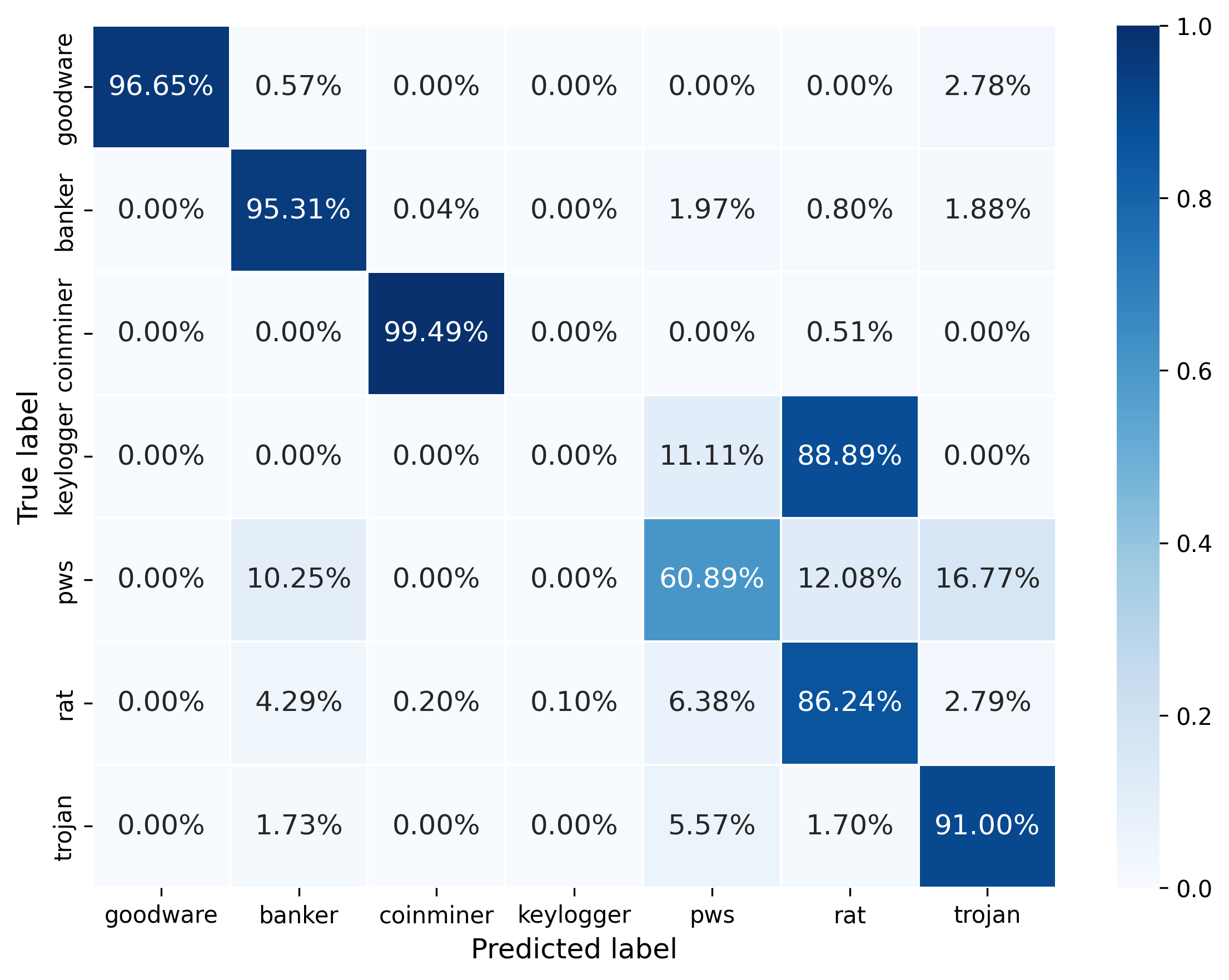}
    \caption{Classification confusion matrix on the raw dataset using the CNN with NLP-based feature engineering.}
    \label{fig:cnn_without_raw}
  \end{subfigure}
  \hfill 
  \begin{subfigure}[b]{0.32\linewidth}
    \includegraphics[width=\linewidth]{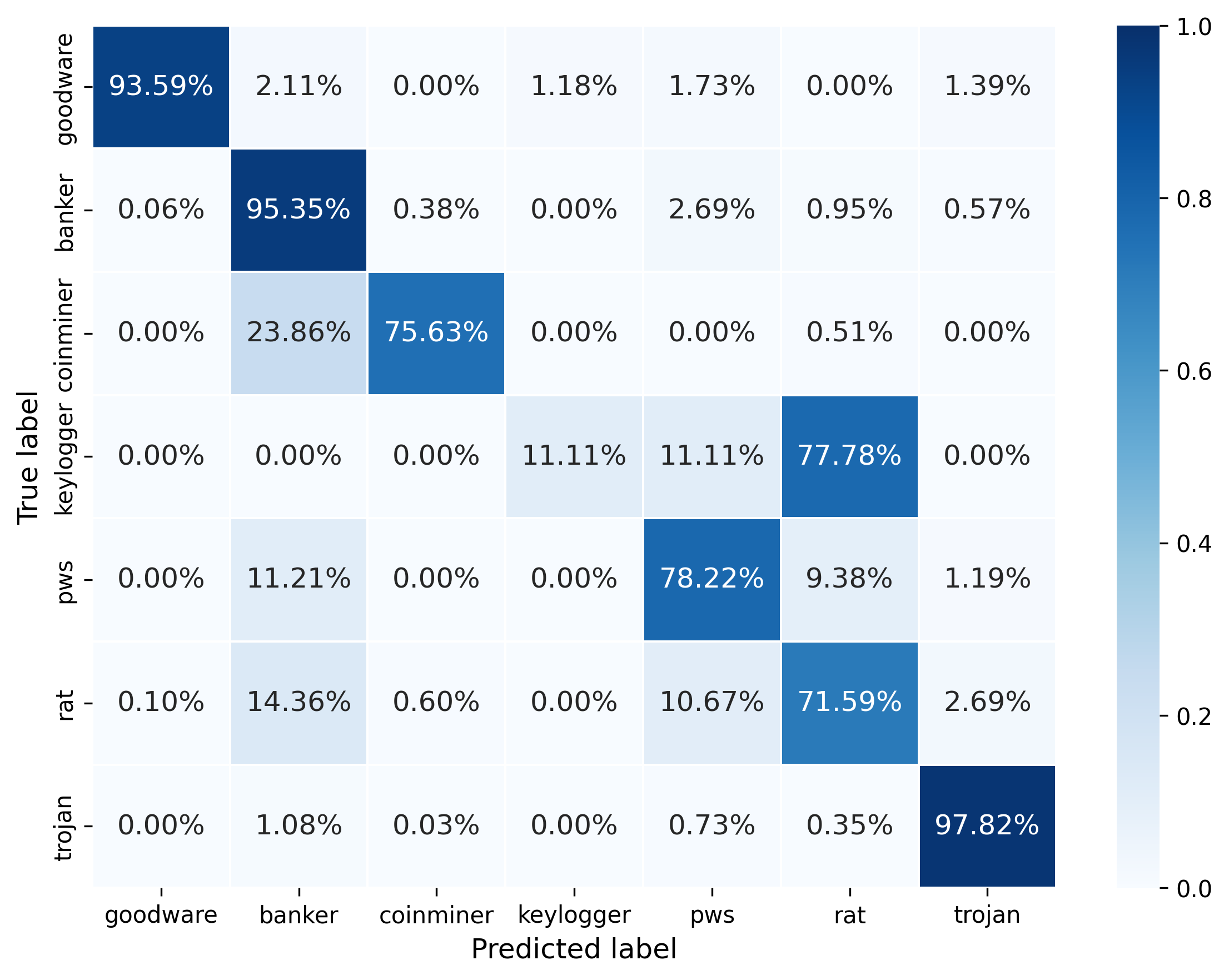}
    \caption{Classification confusion matrix on the raw dataset using the Transformer with NLP-based feature engineering.}
    \label{fig:tran_without_raw}
  \end{subfigure}
  \begin{subfigure}[b]{0.32\linewidth}
    \includegraphics[width=\linewidth]{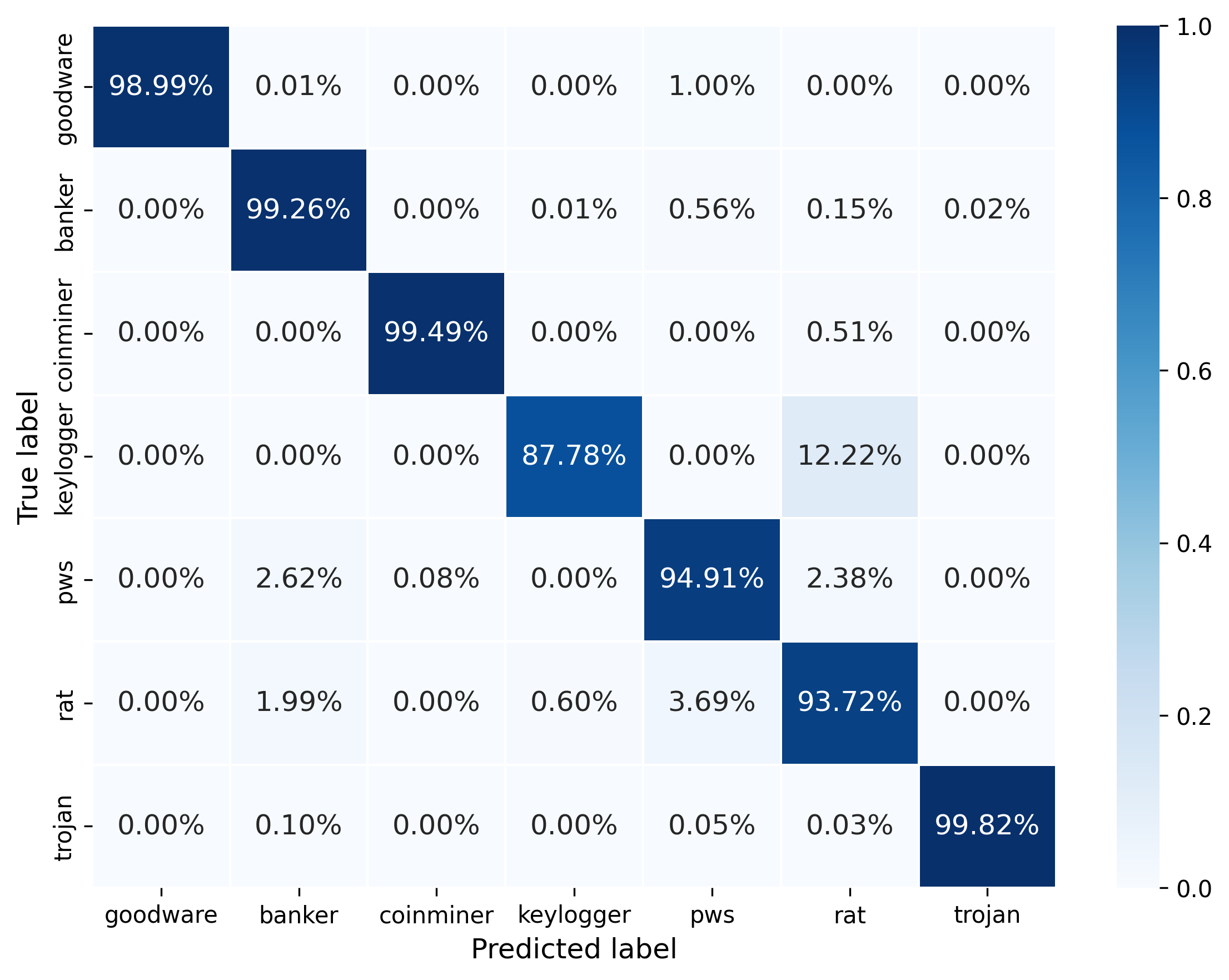}
    \caption{Classification confusion matrix on the supplementary dataset using the CNN with knowledge-based feature engineering.}
    \label{fig:cnn_with_supple}
  \end{subfigure}
  \hfill 
  \begin{subfigure}[b]{0.32\linewidth}
    \includegraphics[width=\linewidth]{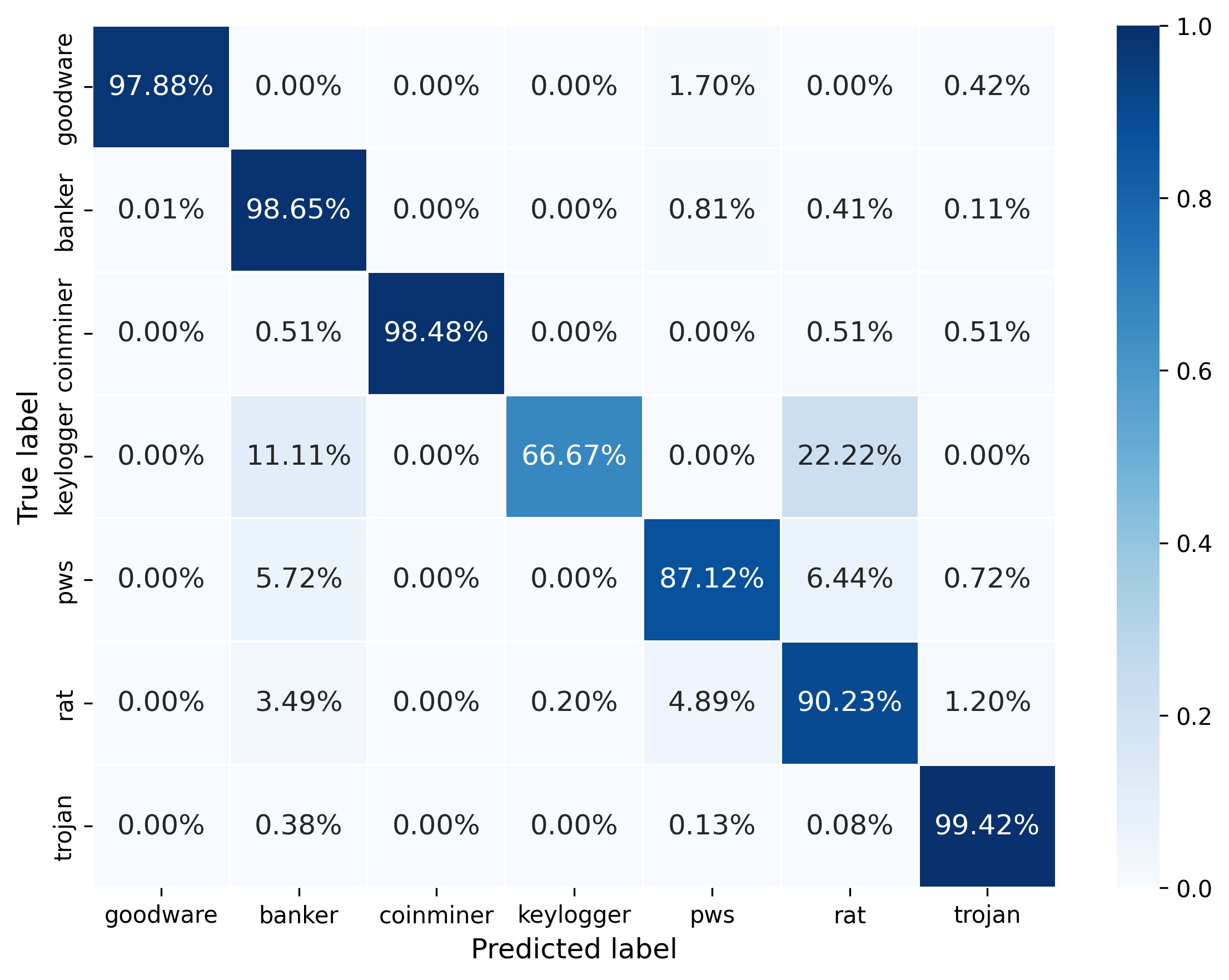}
    \caption{Classification confusion matrix on the supplementary dataset using the CNN with NLP-based feature engineering.}
    \label{fig:cnn_without_supple}
  \end{subfigure}
  \hfill 
  \begin{subfigure}[b]{0.32\linewidth}
    \includegraphics[width=\linewidth]{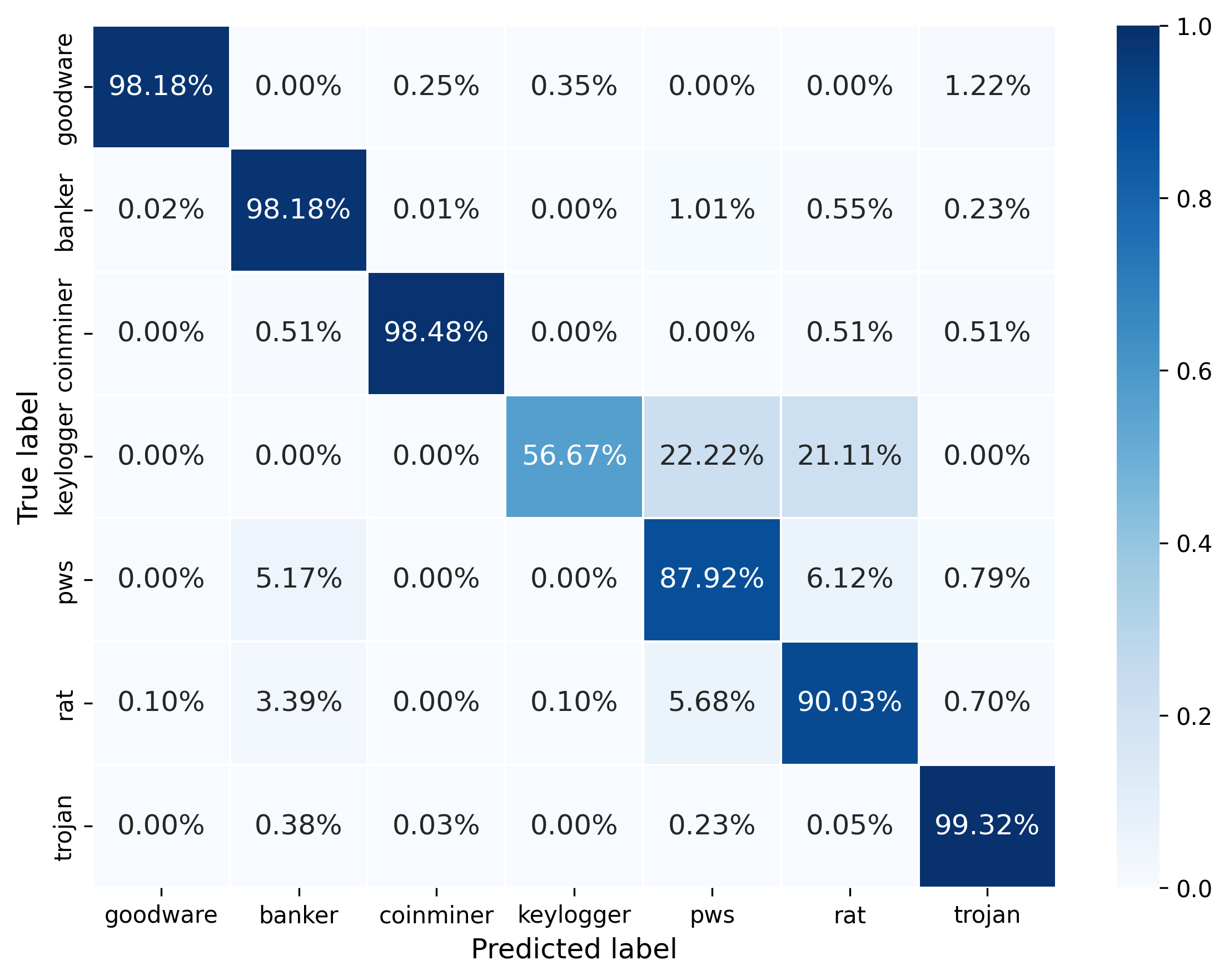}
    \caption{Classification confusion matrix on the supplementary dataset using the Transformer with NLP-based feature engineering.}
    \label{fig:tran_without_supple}
  \end{subfigure}
  
  \caption{Classification confusion matrices using different processing methods on the raw and supplemented datasets.}
  \label{fig:matrice_results}
\end{figure*}

Figs.~\ref{fig:cnn_with_raw} and~\ref{fig:cnn_without_raw} display the classification confusion matrices on the raw dataset, using the CNN model with knowledge-based and NLP-based feature engineering, respectively. The classification accuracy for the "keylogger" category is nearly 0\% in the model with NLP-based feature engineering. This outcome also explains why the CNN model, despite showing high ACC and PRE on the raw dataset, exhibits significantly lower RC and F1 scores. The primary reason for this is the insufficient number of samples in the "keylogger" category, which causes the model fail to effectively learn the characteristics of this category during the training, resulting in poor recognition performance during the testing. However, after applying knowledge-based feature engineering, the dataset shows marked improvement, especially in classifying minority classes, highlighting the significant effect of knowledge-based feature engineering in mitigating data imbalance.

Figs.~\ref{fig:cnn_with_supple} and~\ref{fig:cnn_without_supple} present the model’s performance on the dataset supplemented with keylogger samples. Under NLP-based feature engineering, the classification accuracy for the keylogger category improves significantly, with the overall performance gains primarily stemming from the classification results for this category. However, despite this improvement, its performance remains inferior to the results achieved by knowledge-based feature engineering. Therefore, the model with knowledge-based feature engineering still outperforms the model with NLP-based feature engineering.

As shown in Figs.~\ref{fig:cnn_without_raw} and~\ref{fig:tran_without_raw}, the Transformer model demonstrates greater sensitivity to keylogger samples with NLP-based feature engineering, outperforming the CNN model by successfully identifying some of these samples. However, as shown in Figs.~\ref{fig:cnn_without_supple} and~\ref{fig:tran_without_supple}, after supplementing the dataset, the performance increase of the Transformer model in this category is limited and does not align with the performance of the CNN model.

\begin{mdframed}
\textbf{Answer for RQ1}: Under the same conditions, models with knowledge-based feature engineering outperform those with NLP-based methods, and this effect is more pronounced when the sample size is small.
\end{mdframed}

It is noteworthy that all models tend to misclassify keylogger samples as RAT type. This is because RATs typically include keyboard input capture functionality, which overlaps with the behavioral characteristics of keyloggers, increasing the difficulty for the model to distinguish between the two.
Moreover, RATs exhibit more diverse and complex behaviors than keyloggers, often incorporating actions such as network communication and privilege escalation, in addition to keyboard capture. These rich features enable the model to more accurately identify RATs, thereby reducing the likelihood of misclassifying RAT samples as keyloggers.

\begin{figure*}[t] 
  \centering
  \begin{subfigure}[b]{0.49\linewidth}
    \includegraphics[width=\linewidth]{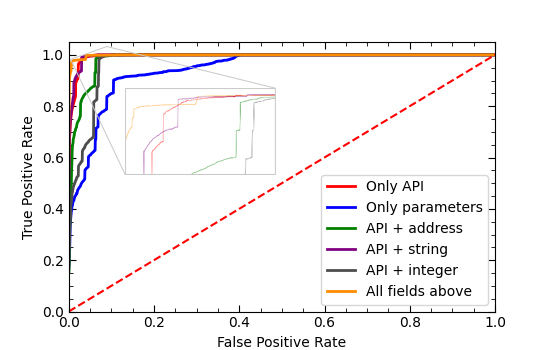}
    \caption{Results of Knowledge-based feature engineering.}
    \label{fig:field_filter_rocs}
  \end{subfigure}
  \hfill 
  \begin{subfigure}[b]{0.49\linewidth}
    \includegraphics[width=\linewidth]{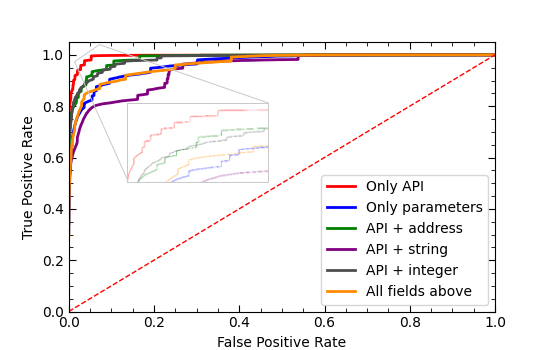}
    \caption{Results of NLP-based feature engineering.}
    \label{fig:filters_no_filters_rocs}
  \end{subfigure}
  \hfill 
 \caption{The impact of different feature combinations on the model's final performance.}
  \label{fig:ablation_results}
\end{figure*}

\subsection{Ablation Study}
\label{sec:ablation_study}
To highlight the utility of various features in malware classification, we examined the impact of different combinations of four types of features on the final performance of the model. For the ablation experiment design, we used the supplementary dataset, as it offers a broad representation of behaviors. In addition to using only API names and only parameters as input features, we also included combinations such as API name with virtual address parameters (API + address), API name with string parameters (API + string), API name with integer parameters (API + integer), and all features together, yielding a total of six combination modes. The performance results are represented using ROC curves, where the True Positive Rate (TPR) indicates the proportion of correctly identified instances of a given class, and the False Positive Rate (FPR) represents the proportion of other classes misclassified as that class. The final ROC curve is generated by averaging the TPR and FPR values across all classes.

As shown in Fig.~\ref{fig:field_filter_rocs}, feature reduction has varying degrees of impact on the model's performance under knowledge-based feature engineering. The performance declines when any single feature is used, compared to using all features. Notably, the combination of API name with string performs similarly to using only the API name feature, while the worst performance is observed when only parameters are used. However, as shown in Fig.~\ref{fig:filters_no_filters_rocs}, under NLP-based feature engineering, using only API name features achieves the best performance, the introduction of any type of parameter feature results in inferior performance compared to using only API name features, with the API + string combination resulting in the lowest performance. We hypothesize that the introduction of numerous string parameters may make it difficult to distinguish whether the input string represents an API name or a parameter, thus disrupting the API patterns in the call sequences. Regardless of the use of feature engineering, the results consistently highlight the critical role of API name in malware classification.

\begin{mdframed}
\textbf{Answer for RQ2}: Regardless of what type of feature engineering is applied, API name features consistently contribute the most to the classification results. After applying knowledge-based feature engineering, incorporating any type of parameter feature leads to an improvement in the model's performance. However, in the NLP-based feature engineering, the performance of the model with any type of parameter features is inferior to that of the model using only API name features.
\end{mdframed}


\section{Key Factor Assessment}

In this section, we employ XAI techniques to assess the key factors behind the model's classification, and further examine artifacts unrelated to a model. In other words, which features of the API sequence contribute the most to the model's classification decision. Based on prior research, a target model can be treated as a black box~\cite{ribeiro2016should,ribeiro2018anchors}. Using simple models to approximate the decision boundaries near the input samples, we identify key features for generating explanations. Moreover, we apply gradient-based methods, utilizing backpropagation in deep neural networks to measure the sensitivity of each feature~\cite{selvaraju2017grad,Lundberg2017}. For Transformer models, we use the attention activation values in the encoder layers to evaluate the relative importance of each token,  providing interpretability analysis.

\subsection{Knowledge-based Method}
Given the use of expert-designed heuristic rules in feature engineering, the model is trained to learn key features. Pearson's correlation coefficient can be used to assess the linear relationship between the extracted features and different malware families. This method does not require understanding the internal structure of the model but rather evaluates the relationship by analyzing the correlation between the features and the output. As shown in Table~\ref{tab:pearson}, the contribution of features varies across different categories, string parameters and address parameters exhibit correlations (r \textgreater~0.6) in distinguishing certain specific categories, such as PWS and goodware, suggesting that these features have a significant impact on predicting the categories and may serve as crucial indicators in the classification process. By contrast, integer parameters are relatively effective in distinguishing goodware, but their overall correlation is low, implying that relying solely on this feature may not accurately differentiate between categories. Notably, virtual addresses exhibit a surprisingly correlation overall.
\begin{table}[t]
\centering
\caption{Pearson correlation coefficients of features.}
\resizebox{0.48\textwidth}{!}{%
\begin{tabular}{ccccccc}
\toprule
\textbf{Type} & \textbf{API Name} & \textbf{API Integer} & \textbf{API String} & \textbf{API Address}  \\
\midrule
Goodware & 0.295 & 0.373 & \textbf{-0.726} & \textbf{-0.483} \\
Banker   & -0.202 & -0.341 & 0.249 & 0.143 \\
Coinminer & -0.39 & 0.018 & -0.042 & \textbf{-0.582}   \\
Keylogger & 0.213 & -0.009 & 0.192 & 0.015  \\
PWS      & 0.06  & 0.096 & \textbf{0.631} & \textbf{-0.422} \\
RAT      & 0.349 & -0.053 & 0.188 & \textbf{0.482} \\
Trojan   & 0.136 & -0.057 & 0.417 & 0.339 \\
All      & \textbf{0.4}   & -0.091 & 0.396 & \textbf{0.512} \\
\bottomrule
\end{tabular}%
}
\label{tab:pearson}
\end{table}
\subsection{NLP-based Method}
In contrast to the manual heuristic rules of knowledge-based feature engineering, the application of NLP methods enables the model to autonomously learn and capture the behavioral features or heuristic patterns. SHAP~\cite{Lundberg2017} is employed to compute the Shapley values of individual tokens, facilitating the differentiation of each feature’s relevance within the malware classification. The token contributions derived here incorporate both positive and negative effects, calculating the cumulative sum of the absolute SHAP values for each token.

As shown in Table~\ref{tab:shaply_token}, in comparison to goodware, most malware types share certain key tokens, such as {\tt modulehandle}, {\tt dll}, and {\tt functionaddress}. These tokens represent fundamental system-level APIs or resource identifiers, which are commonly found across various types of malware. This commonality indicates that underlying system resources are prevalent features in malicious activities. It is noteworthy that the model often relies on handles and virtual addresses in classification tasks, but they are difficult for humans to interpret and vary across different executions of the same binary. This makes them unsuitable as stable indicative features for classification. Methods such as feature hashing~\cite{weinberger2009feature} are needed to improve their generalizability, which highlights the importance of manual feature engineering.

Time-related tokens, such as \texttt{milliseconds} and \texttt{NtDelayExecution}, exhibit significant weights across various malware types, particularly within the RAT category. This suggests that malware commonly utilizes timing-based scheduling and delay execution strategies to evade detection, thereby enhancing the stealthiness of malicious activities.

\begin{table}[t!]
\centering
\caption{The top 10 individual tokens influencing the model's decision across different malware types.}
\begin{tabular}{l@{\hspace{10pt}}l}
    \toprule
    \textbf{Types} & \textbf{Top 10 individual Tokens (SHAP Value)} \\ 
    \midrule
    Goodware & \begin{tabular}[t]{@{}l@{}}hkey, lpsubkey, regopenkeyexw, local, \\ machine, software, microsoft, hfile, \\ 1, generic\end{tabular} \\
    \midrule
    Banker & \begin{tabular}[t]{@{}l@{}}\textbf{modulehandle}, 0, ldrgetprocedureaddress, \\ ordinal, \textbf{functionaddress}, \textbf{dll}, \\ 0x00000000, kernel32, baseaddress, \\ 0xffffffff\end{tabular} \\
    \midrule
   
    Coinminer & \begin{tabular}[t]{@{}l@{}}kernel32, disableusermodecallbackfilter,\\
    \textbf{dll}, getnumberofconsoleinputevents, \\
    \textbf{modulehandle}, \textbf{functionaddress}, \\
    ldrgetprocedureaddress, \\ 0xfffffffe, 0, getsystemtimeasfiletime\end{tabular} \\
    \midrule
     Keylogger & \begin{tabular}[t]{@{}l@{}}\textbf{functionaddress}, regqueryvalueexw, \\
     ordinal, ntdelayexecution, \textbf{modulehandle}, \\
     \textbf{dll}, milliseconds, kernel32, 0xffffffff, \\ stackpivoted\end{tabular} \\
    \midrule
    PWS & \begin{tabular}[t]{@{}l@{}}disableusermodecallbackfilter, \texttt{<unk>}, \\ \textbf{modulehandle}, \textbf{dll}, oleaut32, 0x00000000, \\ getsystemtimeasfiletime, \textbf{functionaddress}, \\ ldrgetprocedureaddress, milliseconds\end{tabular} \\
    \midrule
    RAT & \begin{tabular}[t]{@{}l@{}}disableusermodecallbackfilter, \\ getsystemtimeasfiletime, \\ \textbf{dll}, \textbf{modulehandle}, 0, \textbf{functionaddress}, \\ ldrgetprocedureaddress, ntdelayexecution,\\ milliseconds, pi\end{tabular} \\
    \midrule
    Trojan & \begin{tabular}[t]{@{}l@{}}disableusermodecallbackfilter, 0x00006000, \\ \textbf{dll}, getsystemtimeasfiletime, oleaut32, 0, \\ milliseconds, \textbf{modulehandle}, ordinal, \\ \textbf{functionaddress}\end{tabular} \\
    \bottomrule
\end{tabular}
\label{tab:shaply_token}
\end{table}

\begin{figure}[t]
  \centering
  \includegraphics[width=0.49\textwidth]{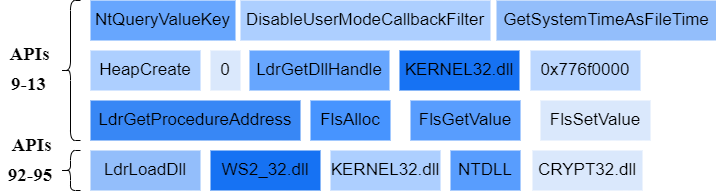}
  \caption{Highlight the areas with the highest attention in the sample under knowledge-based feature engineering processing, where darker colors indicate greater attention.}
  \label{fig:active_feature}
\end{figure}
\subsection{The Attention Mechanism}
For the Transformer model, attention activations can be used to indicate the relative importance of different tokens within the model. We present an analysis of a sample classified as Trojan (MD5:\textit{ 227958c8e6e50ac28ffeb146156e82a5}), focusing on the biases in the attention weights and their implications.

In the knowledge-based feature engineering approach, the model particularly highlights two segments of function calls in the sample sequence: between tokens 9-13 and 92-98, as shown in Fig.~\ref{fig:active_feature}. In the first group of API calls, \texttt{NtQueryValueKey} is a system call used for querying Windows registry values. The model focuses specifically on the query of the registry key \texttt{DisableUserModeCallbackFilter}, which is commonly associated with the anti-debugging or virtual machine detection. Additionally, \texttt{LdrGetDllHandle} is used to retrieve the handle of a loaded DLL, and \texttt{LdrGetProcedureAddress} is used to retrieve the address of an exported function from a DLL. These API calls are commonly involved in dynamically loading libraries and accessing specific functions, which may be a part of the process of setting up or interacting with system components, and potentially installing system hooks to capture user operations.

\begin{figure}[t]
  \centering
  \includegraphics[width=0.5\textwidth]{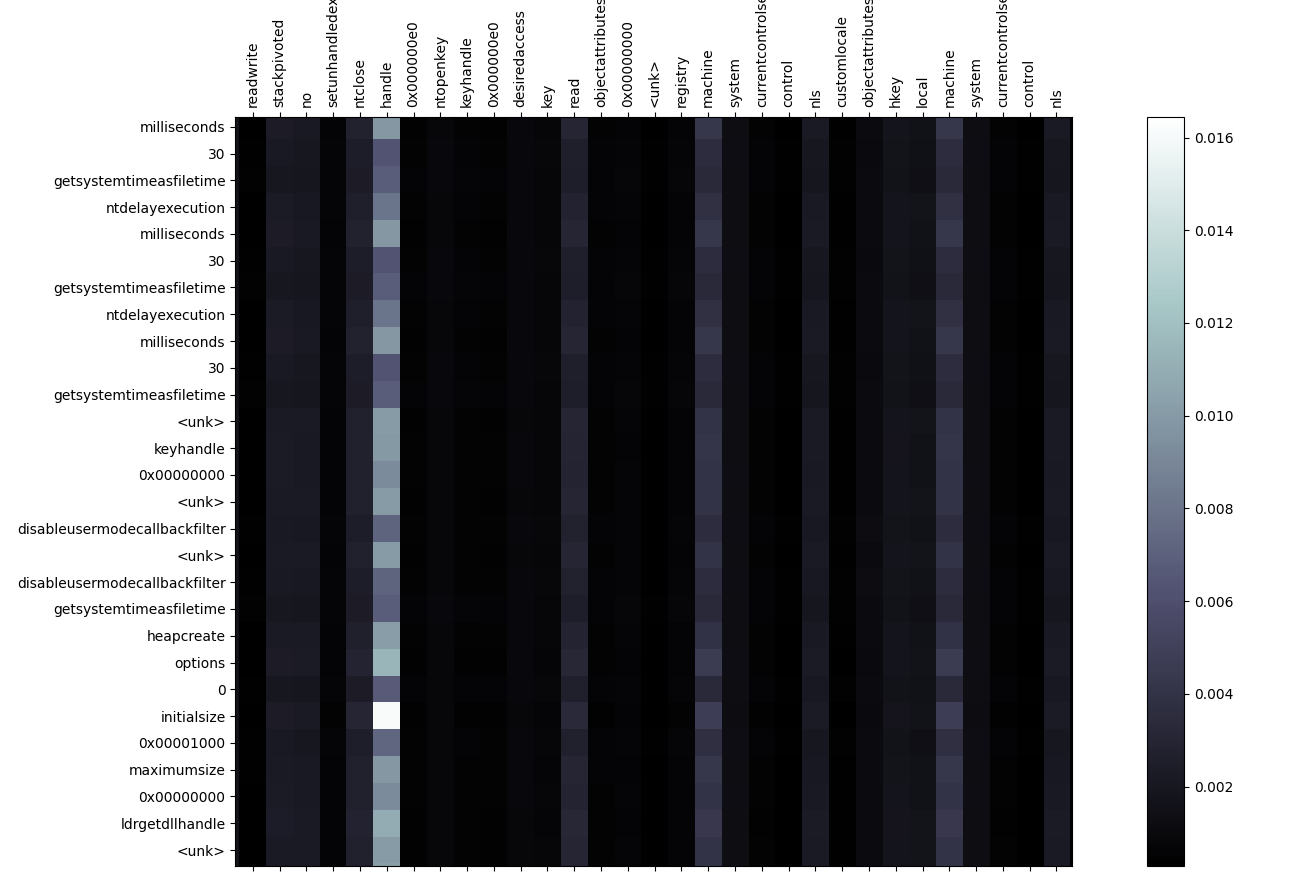}
  \caption{Attention activations for different tokens in the first attention layer of the model.}
  \label{fig:active_without}
  \vspace{-0.1in}
\end{figure}

As shown in Fig.~\ref{fig:active_without}, by contrast, the model with NLP-based feature engineering primarily focuses on APIs related to system time retrieval (\texttt{GetSystemTimeAsFileTime}), delayed execution (\texttt{NtDelayExecution}), and registry access (\texttt{keyhandle}). These actions mainly reflect the anti-debugging behaviors; in particular, the delayed execution is typically used to outlast the analysis periods, assuming that the analysis lasts only for a limited time. However, these anti-debugging and anti-sandbox techniques are commonly used in malware and are not those core characteristics of Trojans. Clearly, their importance is much less than that of behaviors, such as installing system hooks, which are emphasized under feature engineering. Notably, both models focus on the \texttt{GetSystemTimeAsFileTime} system call, which retrieves and returns the current system time as a file time in the timestamp format. This function is commonly used for logging or event time synchronization. Malware may leverage this to mark the time of its operations or synchronize events during the execution of malicious activities.

\begin{mdframed}
\textbf{Answer for RQ3}: A model typically relies on API names, handles, and library addresses to distinguish between different types of malware. Models with NLP-based feature engineering tend to focus more on superficial features, such as anti-debugging behaviors, while models with knowledge-based feature engineering are more effective at capturing the core characteristics of samples (e.g., the malicious behaviors like installing system hooks).
\end{mdframed}

\section{Discussion}
\label{sec:intro_discu}
\textit{1) Threat to validity:} Like all empirical studies, our observations are subject to some threats to the validity. In the following, we discuss the potential sources of bias and how we address them with our best-efforts. 

To ensure that those variations in model performance are solely attributed to different feature engineering approaches rather than other external factors, we maintain consistency in data sources and preserve uniformity in computational methods and evaluation criteria. To mitigate the selection bias, we carefully design inclusion and exclusion criteria for the dataset to achieve a sample balance. Additionally, to reduce the impact of the model selection bias, we conduct our experiments using three models with distinct characteristics, thereby mitigating the result deviations caused by the differences in model architectures. 
Regarding the external validity threats, particularly the generalizability of our findings, we acknowledge that the applicability of this study is constrained by the analyzed data sources. 

Given the widespread adoption of API analysis in the security research, our work primarily focuses on different feature engineering techniques for API call sequences. We believe our findings hold a certain degree of representativeness in this domain. However, since model performance may vary with the changes in the dataset characteristics, a careful evaluation is required when applying our conclusions to other data environments.

\textit{2) Recommendations:} The models using knowledge-based feature engineering outperform the models using NLP-based methods. Due to the disparity between the features utilized by ML models and those selected by human analysts, it is necessary to reassess the potential features that are often overlooked due to their interpretability challenges for humans. When introducing new features, caution must be exercised in organizing the feature inputs to avoid interference or redundancy among features.

API name features play a critical role in malware classification. Given the significant pattern features present in API calls, especially local API call patterns, the model design should prioritize the architectures such as CNN that are more focused on capturing local features, to improve classification accuracy.

  Models often rely on fixed-value features in classification tasks, such as handles and address values. Such a dependency may limit the model’s generalization ability, as it could become overly dependent on these fixed-value features, and thus reducing its capability to identify broader patterns of malicious behaviors. These features require manual feature engineering to be effectively utilized.

\textit{3) Limitation:}
When being executed in virtualized or emulated environments, some malware may avoid exposing their malicious behaviors due to the intervention of environment-aware detection mechanisms~\cite{li2020adversarial}. The dataset we used does not account for such a situation. Malware that focuses on sandbox evasion techniques~\cite{lindorfer2011detecting} does not produce high-quality execution reports, making it difficult to determine whether it is malware. This highlights the need for techniques to counter evasion tactics in production environments.

\section{Conclusion}
\label{sec:intro_conclu}
In this paper, we provide new insights into the impact of feature engineering upon malware classification. We used the complete set of data features from API call sequences, applying both knowledge-based and NLP-based feature engineering methods and evaluated their performance under three ML models. Our results show that knowledge-based feature engineering methods consistently outperform NLP-based ones, particularly when dealing with small sample datasets. Regardless of the feature engineering method being applied, CNN consistently achieves the best performance among three ML models. In addition, there is a stark contrast in the performance of knowledge-based and NLP-based methods when increasing the number of input features. Our analysis further reveals that models often focus on features such as handles and virtual address values, which are challenging for humans to interpret and  also execution-environment dependent.


\section{Acknowledgements}
We thank the anonymous reviewers for their insightful and constructive feedback. This work was supported by Network Twinning Enhancement Methods project (No.E3Z011104) of Climbing Plan from Institute of Information Engineering, Chinese Academy of Science, and by the Industrial Foundation Reconstruction and High-Quality Development of Manufacturing Industry Special Project (0747-2361SCCZA193). Zhi Li is the corresponding author.

\bibliographystyle{IEEEtran}   
\bibliography{main}

\pagebreak

\appendix

The table below shows the optimal hyperparameters for different models to reproduction.

\begin{table}[H]
\centering
\caption{Key Hyperparameters and Optimal Values for the Models}
\label{tab:hyperparameters}
\renewcommand{\arraystretch}{1}  
\small
\begin{tabular}{l@{\hspace{4pt}}l@{\hspace{4pt}}c@{\hspace{4pt}}c}
\hline
\textbf{Model} & \textbf{Hyperparameter}  & \multicolumn{2}{c}{\textbf{Optimal Value}}  \\
\cline{3-4}
&  & \textbf{Knowledge} & \textbf{NLP} \\
\hline
\multirow{4}{*}{\centering CNN}  & Vector dimensions & 132 & 96 \\ 
  & k-gram  & 2, 3, 4, 5 & 2, 3, 4, 5 \\
  & Kernel channel   & 128 & 128 \\
  & Dropout rate & 0.3 & 0.3 \\
\hline
\multirow{4}{*}{\centering LSTM} & Vector dimensions & 132 & 64 \\ 
  & LSTM hidden units  & 64 & 256 \\
  & LSTM layers  & 2 & 1 \\
  & LSTM dropout  & 0.1 & 0.1 \\
\hline
\multirow{5}{*}{\centering Transformer} & Vector dimensions & 132 & 64 \\
  & Number of attention heads  & 8 & 8 \\
  & Hidden layer dimensions  & 256 & 256 \\
  & Number of encoder layers &  2 & 2 \\
  & Dropout & 0.2 & 0.3 \\
\hline
\multirow{2}{*}{\centering MLP}  & Units of hidden layers & (128, 64) & (128, 64) \\
  & Dropout rate & 0.2 & 0.2 \\
\hline
\end{tabular}
\end{table}

\end{document}